\documentclass[aip,reprint,graphicx]{revtex4-1}
\usepackage{graphicx}

\draft 

\begin{document}


\title{Simultaneous modulation transfer spectroscopy on transitions of multiple atomic species for compact laser frequency reference modules} 



\author{Moritz Mihm}
\email[]{mmihm@uni-mainz.de}
\author{Kai Lampmann}
\author{Andr\'e Wenzlawski}
\author{Patrick Windpassinger}
\affiliation{Johannes Gutenberg-Universit\"at Mainz, Staudingerweg 7, 55128 Mainz}


\date{\today}

\begin{abstract}
We present a technique for simultaneous laser frequency stabilization on transitions of multiple atomic species with a single optical setup. The method is based on modulation transfer spectroscopy and the signals are separated by modulating at different frequencies and electronically filtered. As a proof of concept, we demonstrate simultaneous spectroscopy of the potassium D$_1$, D$_2$ and rubidium D$_2$ transitions. The technique can easily be extended to other atomic species and allows the development of versatile and compact frequency reference modules.

\end{abstract}


\maketitle 

\section{Introduction}
Lasers, referenced to a well defined frequency like that of an atomic transition or an ultra-stable optical resonator, are key elements in a variety of quantum technology applications. For instance, atom interferometers \cite{Kasevich1991}, optical clocks \cite{Ludlow2015} or ion based quantum computers \cite{Haffner2008} require a number of frequency stabilized lasers, e.g. for laser cooling and trapping or state interrogation. While techniques for stabilizing lasers to atomic vapor samples are in general well established \cite{Drever1983, Bjorklund1983} and used in atomic physics labs around the world, the field application and further exploitation of aforementioned quantum technologies requires miniaturized, compact and robust technical realizations. One key aspect in this context is e.g. the reduction of complexity to reduce the probability of failure, and the reduction of components to decrease the system size and to limit costs.

One main issue in connection with system size, cost and complexity is that especially in the case of multi species experiments, several optical spectroscopy setups which require independent alignment are run in parallel, i.e. individual setups for every atomic species and every atomic transition frequency are usually implemented. To counter this approach, we have as a proof of concept implemented simultaneous dual species (potassium, rubidium), multi transition line (D$_1$, D$_2$), single beam modulation transfer spectroscopy in a single dual species atomic vapor cell for three separate lasers. By multi frequency modulation/demodulation in a Pound-Drever-Hall type \cite{Drever1983} setup for each laser, we isolate the spectroscopy signals of the different sources electronically after detection on a single photodiode and generate the individual laser's error signal. The number of involved optical components is thereby reduced significantly compared to three stand-alone setups. This approach can easily be extended to more lasers and more atomic species according to the needs of the respective application.

\section{Experimental setup}
\begin{figure*}
	\includegraphics[width=15cm]{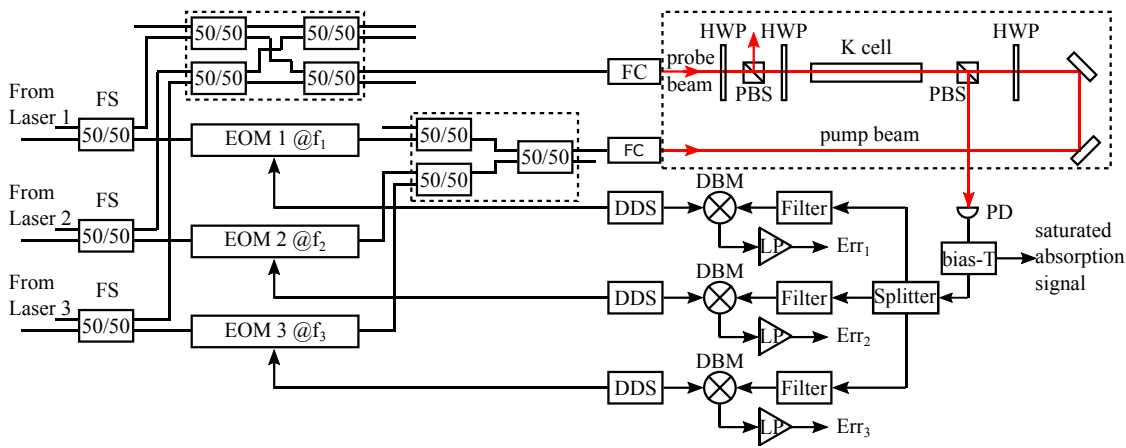}%
	\caption{Schematic setup for simultaneous modulation transfer spectroscopy of the potassium D$_1$, D$_2$ and rubidium D$_2$ transitions. DBM: double-balanced mixer; DDS: direct digital synthesis; EOM: electro-optic modulator; FC: fiber collimator; FS: fiber splitter; 50/50: splitting ratio; HWP: half-wave plate; LP: low-pass filter; PBS: polarizing beam splitter; PD: photodiode.\label{fig:setup}}%
\end{figure*}

To generate an error signal, we employ modulation transfer spectroscopy, the overall approach would however work just as well with standard frequency modulation spectroscopy. Modulation transfer spectroscopy has been discussed individually for potassium and rubidium \cite{Mudarikwa2012, Noh2011, Zhang2003, McCarron2008}. In order to excite the different optical transitions simultaneously, three lasers at wavelengths of 770.1\,nm ($^{39}$K D$_1$), 766.7\,nm ($^{39}$K D$_2$) and 780.2\,nm ($^{85, 87}$Rb D$_2$) are required. To overlap the optical paths and to induce the required phase modulation of the laser light, we use a fiber-based system. This also keeps the experiment clear and compact. A general schematic of our setup is depicted in Fig. \ref{fig:setup}.

The light of the three lasers (two home-built DFB lasers with diodes by eagleyard Photonics and one\linebreak TOPTICA Photonics DL pro) is initially coupled into polarization maintaining fibers. Separation of pump and probe beam is achieved by 2x2 fiber splitters\linebreak (Evanescent Optics 954P). One output port of each of the three splitters is directly connected to another splitter (4x4) for combination of all probe light. Each of the other ports is connected to a fiber electro-optic phase modulator (EOM; Photline NIR-MPX800) for frequency sideband generation. The modulated beams are overlapped with another splitter (4x2) to generate the combined pump light for the spectroscopy. With this arrangement, the power ratios between pump and probe ex fiber are equal, apart from the individual losses of the EOMs. However, the power ratios can be optimized with conventional free space optics.

To separate the photodetector signals of the three lasers, the EOMs are driven at modulation frequencies of 4.5\,MHz, 11.5\,MHz and 16\,MHz. This takes advantage of standard Mini-Circuits bandpass filter windows between $\approx$7\,MHz and $\approx$15\,MHz and the frequencies are chosen such that they are no multiples of each other. The frequencies are generated by three dual-channel direct digital synthesis (DDS) based frequency generators (Rigol DG1022) which allow us to adjust the phase between the local oscillator (LO) and the spectroscopy signal from the radio-frequency (RF) photodiode. The collimated beams for pump and probe are guided through a quartz spectroscopy cell with conventional free space optics in a counter-propagating fashion. The vapor cell (Sacher Lasertechnik VC-K-09-50-Q) is intended for potassium spectroscopy but also contains atomic rubidium. A home-built heating around the cell is used to increase the vapor pressure and hereby the signal strength. The cell is heated to temperatures around $65^\circ$C at the windows.

The probe beam is detected on a silicon photodiode with 150\,MHz bandwidth (Thorlabs PDA10A-EC). The low- and high-frequency components of the photodiode signal are separated using a bias-T (Mini-Circuits\linebreak ZFBT-6GW+). The low-frequency component represents the saturated absorption spectrum and is directly monitored. The high-frequency component is split using a three way power splitter\linebreak (Mini-Circuits ZFSC-3-1W-S+) and electronically separated by low, high and bandpass filters (Mini-Circuits BLP-5+, SHP-20+, SBP-10.7+) according to the modulation frequencies. The signal of each channel is amplified (Mini-Circuits\linebreak ZFL-500+), mixed down (Mini-Circuits ZFM-3-S+) with the corresponding local oscillator and filtered using a low pass filter with cutoff frequency 1.9\,MHz (Mini-Circuits SLP-1.9+).

\section{Experimental results}
\begin{figure}
	\includegraphics[width=8.5cm]{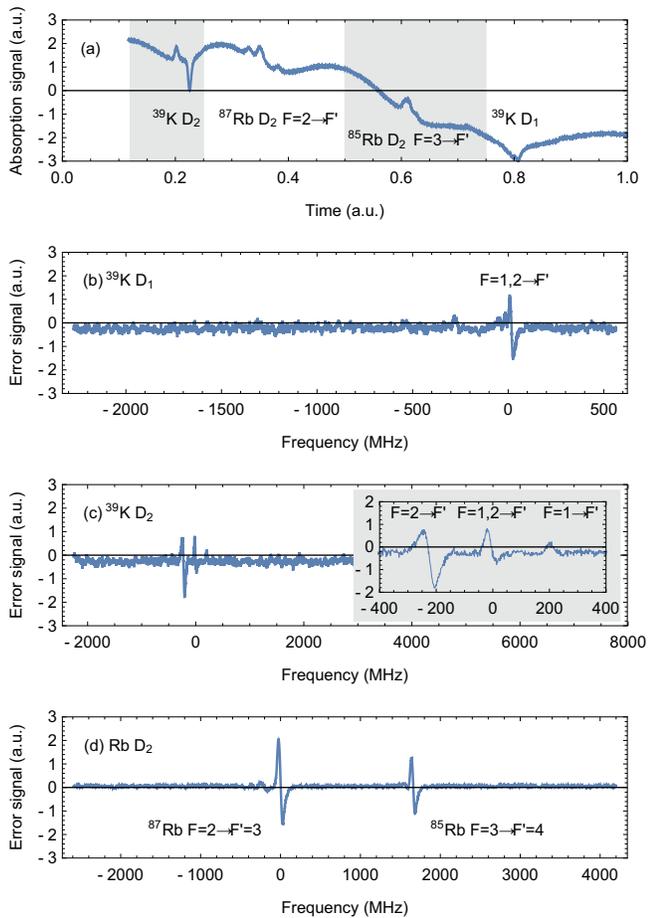}
	\caption{Saturated absorption spectrum and associated error signals of the potassium D$_1$, D$_2$ and rubidium D$_2$ transitions.\label{fig:spectra}}%
\end{figure}

To show the capabilities of the setup, we frequency scan the three lasers simultaneously across the atomic transitions in question. When we scan with $\approx$470\,MHz/ms (K D$_1$), $\approx$1.7\,GHz/ms (K D$_2$) and $\approx$1.1\,GHz/ms (Rb D$_2$) in an appropriate range, we observe the saturated absorption spectrum as displayed in Fig. \ref{fig:spectra} (a). As indicated in the plot, the spectrum is an overlap of three individual saturated absorption spectra.

In our setup, the relative pump and probe powers of the different lasers ex fiber are fixed due to the chosen fiber splitter ratios. The power ratios are optimized with conventional free space optics. These powers are in general a trade-off between power broadening and signal-to-noise ratio of the associated error signals. For the chosen trade-off, the D$_1$ and D$_2$ transitions for the most abundant naturally occurring isotope of potassium ($^{39}$K) can easily be identified in the saturated absorption spectrum. For the D$_2$ transition of rubidium, we set the scan range such that only the $F=2\rightarrow F'$ of $^{87}$Rb and $F=3\rightarrow F'$ of $^{85}$Rb set of transitions are displayed to avoid confusion. We use the notation $F=x\rightarrow F'$ for the set of transitions from state $x$ in the manifold of excited states.

With optimized phases of the respective local oscillators at the mixers in the high-frequency components of the photodetector signal, we obtain the desired individual error signal for each of the three lasers separately as shown in Fig. \ref{fig:spectra} (b)-(d). The error signals' x-axes are frequency calibrated with the respective absorption spectrum. The main peak in the error signal of potassium D$_1$ is the crossover transition $F=1,2\rightarrow F'$. As part of the calibration, this transition is used as zero-frequency reference. The error signal of potassium D$_2$ shows three peaks which can be identified as the set of ground state crossover transitions $F=1,2\rightarrow F'$ (at zero) as well as the set of transitions $F=2\rightarrow F'$ (left peak) and $F=1\rightarrow F'$ (right peak). The two peaks in the rubidium error signal stem from the closed transition $F=2\rightarrow F'=3$ of $^{87}$Rb (left peak at zero) and the closed transition $F=3\rightarrow F'=4$ of $^{85}$Rb (right peak). These signals can now be used for frequency stabilization with standard servo amplifiers.

By temporarily blocking the light from two lasers at a time, we can directly compare the error signals of single and multi species spectroscopy for each species. Taking the peak-to-peak amplitude of the largest signal divided by the standard deviation of the noise floor as measure for the signal-to-noise ratio (SNR), we achieve the values summarized in Table \ref{tab:snr}.

\begin{table}
\caption{Error signal signal-to-noise ratios (SNR) of single and multi species spectroscopy for each species.\label{tab:snr}}
\begin{ruledtabular}
\begin{tabular}{lccc}
	SNR							& K D$_1$	& K D$_2$	& Rb D$_2$\\\hline
	Single species spectroscopy	& 33			& 69			& 109\\
	Multi species spectroscopy	& 28			& 59			& 121\\
\end{tabular}
\end{ruledtabular}
\end{table}

\section{Conclusion}
We successfully demonstrated a technique for simultaneous modulation transfer spectroscopy of multiple atomic species. Isolating the individual signals by modulating at different frequencies and electronic filtering, we were able to generate three error signals at the potassium D$_1$, D$_2$ and rubidium D$_2$ transitions. The obtained signals are  comparable to former frequency reference modules \cite{Lezius2016, Dinkelaker2017, Schkolnik2016} used for different space missions. Our concept is especially applicable in dual species experiments using potassium and rubidium atoms \cite{Kulas2017, Menoret2011}, taking new hybrid cooling schemes \cite{Nath2013, Salomon2013, Chen2016} into account by including the potassium D$_1$ transition. Our technique can easily be extended to other atomic species.

At the same time, our technique only requires a single optical setup saving optical components and weight, and reducing complexity. This allows the development of compact frequency reference modules for experiments with multiple atomic species. The approach is for instance ideally suited for an implementation in highly stable ZERODUR\textregistered{} based optical benches \cite{Duncker2014} for space or other field applications.


%
%

%

\begin{acknowledgments}
Our work is supported by the German Space Agency DLR with funds provided by the Federal Ministry for Economic Affairs and Energy (BMWi) under grant numbers 50 WP 1433 and 50 WP 1703.
\end{acknowledgments}


\begin{thebibliography}{10}
	
	\bibitem{Kasevich1991}
	Mark Kasevich and Steven Chu.
	\newblock Atomic interferometry using stimulated raman transitions.
	\newblock {\em Phys. Rev. Lett.}, 67:181--184, Jul 1991.
	
	\bibitem{Ludlow2015}
	Andrew~D. Ludlow, Martin~M. Boyd, Jun Ye, E.~Peik, and P.~O. Schmidt.
	\newblock Optical atomic clocks.
	\newblock {\em Rev. Mod. Phys.}, 87:637--701, Jun 2015.
	
	\bibitem{Haffner2008}
	H.~H\"affner, C.F. Roos, and R.~Blatt.
	\newblock Quantum computing with trapped ions.
	\newblock {\em Physics Reports}, 469(4):155 -- 203, 2008.
	
	\bibitem{Drever1983}
	R.~W.~P. Drever, J.~L. Hall, F.~V. Kowalski, J.~Hough, G.~M. Ford, A.~J.
	Munley, and H.~Ward.
	\newblock Laser phase and frequency stabilization using an optical resonator.
	\newblock {\em Applied Physics B}, 31(2):97--105, Jun 1983.
	
	\bibitem{Bjorklund1983}
	G.~C. Bjorklund, M.~D. Levenson, W.~Lenth, and C.~Ortiz.
	\newblock {Frequency modulation (FM) spectroscopy}.
	\newblock {\em Applied Physics B Photophysics and Laser Chemistry},
	32(3):145--152, 1983.
	
	\bibitem{Mudarikwa2012}
	L~Mudarikwa, K~Pahwa, and J~Goldwin.
	\newblock {Sub-Doppler modulation spectroscopy of potassium for laser
		stabilization}.
	\newblock {\em Journal of Physics B: Atomic, Molecular and Optical Physics},
	45(6):065002, 2012.
	
	\bibitem{Noh2011}
	Heung-ryoul Noh, Sang~Eon Park, Long~Zhe Li, and Chang-ho Cho.
	\newblock {Rb atoms : theory and experiment}.
	\newblock {\em Optics Express}, 19(23):490--492, 2011.
	
	\bibitem{Zhang2003}
	Jing Zhang, Dong Wei, Changde Xie, and Kunchi Peng.
	\newblock {Characteristics of absorption and dispersion for rubidium D$_2$
		lines with the modulation transfer spectrum}.
	\newblock {\em Optics Express}, 11(11):1338, 2003.
	
	\bibitem{McCarron2008}
	D.~J. McCarron, S.~A. King, and S.~L. Cornish.
	\newblock {Modulation transfer spectroscopy in atomic rubidium}.
	\newblock {\em Measurement Science and Technology}, 19(10), 2008.
	
	\bibitem{Lezius2016}
	Matthias Lezius, Tobias Wilken, Christian Deutsch, Michele Giunta, Olaf Mandel,
	Andy Thaller, Vladimir Schkolnik, Max Schiemangk, Aline Dinkelaker, Anja
	Kohfeldt, Andreas Wicht, Markus Krutzik, Achim Peters, Ortwin Hellmig, Hannes
	Duncker, Klaus Sengstock, Patrick Windpassinger, Kai Lampmann, Thomas
	H{\"{u}}lsing, Theodor~W. H{\"{a}}nsch, and Ronald Holzwarth.
	\newblock {Space-borne frequency comb metrology}.
	\newblock {\em Optica}, 3(12):1381, 2016.
	
	\bibitem{Dinkelaker2017}
	Aline~N. Dinkelaker, Max Schiemangk, Vladimir Schkolnik, Andrew Kenyon, Kai
	Lampmann, Andr{\'{e}} Wenzlawski, Patrick Windpassinger, Ortwin Hellmig,
	Thijs Wendrich, Ernst~M. Rasel, Michele Giunta, Christian Deutsch, Christian
	K{\"{u}}rbis, Robert Smol, Andreas Wicht, Markus Krutzik, and Achim Peters.
	\newblock {Autonomous frequency stabilization of two extended-cavity diode
		lasers at the potassium wavelength on a sounding rocket}.
	\newblock {\em Applied Optics}, 56(5):1388--1396, 2017.
	
	\bibitem{Schkolnik2016}
	V.~Schkolnik, O.~Hellmig, A.~Wenzlawski, J.~Grosse, A.~Kohfeldt,
	K.~D{\"{o}}ringshoff, A.~Wicht, P.~Windpassinger, K.~Sengstock, C.~Braxmaier,
	M.~Krutzik, and A.~Peters.
	\newblock {A compact and robust diode laser system for atom interferometry on a
		sounding rocket}.
	\newblock {\em Applied Physics B}, 2016.
	
	\bibitem{Kulas2017}
	Sascha Kulas, Christian Vogt, Andreas Resch, Jonas Hartwig, Sven Ganske, Jonas
	Matthias, Dennis Schlippert, Thijs Wendrich, Wolfgang Ertmer, Ernst {Maria
		Rasel}, Marcin Damjanic, Peter We{\ss}els, Anja Kohfeldt, Erdenetsetseg
	Luvsandamdin, Max Schiemangk, Christoph Grzeschik, Markus Krutzik, Andreas
	Wicht, Achim Peters, Sven Herrmann, and Claus L{\"{a}}mmerzahl.
	\newblock {Miniaturized Lab System for Future Cold Atom Experiments in
		Microgravity}.
	\newblock {\em Microgravity Science and Technology}, 29(1-2):37--48, 2017.
	
	\bibitem{Menoret2011}
	V.~M{\'{e}}noret, R.~Geiger, G.~Stern, N.~Zahzam, B.~Battelier, A.~Bresson,
	A.~Landragin, and P.~Bouyer.
	\newblock {Dual-wavelength laser source for onboard atom interferometry}.
	\newblock {\em Optics Letters}, 36(21):4128, 2011.
	
	\bibitem{Nath2013}
	Dipankar Nath, R.~Kollengode Easwaran, G.~Rajalakshmi, and C.~S. Unnikrishnan.
	\newblock {Quantum-interference-enhanced deep sub-Doppler cooling of 39K atoms
		in gray molasses}.
	\newblock {\em Physical Review A - Atomic, Molecular, and Optical Physics},
	88(5):1--7, 2013.
	
	\bibitem{Salomon2013}
	G.~Salomon, L.~Fouch{\'{e}}, P.~Wang, A.~Aspect, P.~Bouyer, and T.~Bourdel.
	\newblock Gray molasses cooling of {$^{39}$}k to a high phase-space density.
	\newblock {\em EPL (Europhysics Letters)}, 104(6):63002, 2013.
	
	\bibitem{Chen2016}
	Hao~Ze Chen, Xing~Can Yao, Yu~Ping Wu, Xiang~Pei Liu, Xiao~Qiong Wang, Yu~Xuan
	Wang, Yu~Ao Chen, and Jian~Wei Pan.
	\newblock {Production of large $^{41}$K Bose-Einstein condensates using $D_1$
		gray molasses}.
	\newblock {\em Physical Review A}, 94(3):1--6, 2016.
	
	\bibitem{Duncker2014}
	Hannes Duncker, Ortwin Hellmig, Andr{\'{e}} Wenzlawski, Alexander Grote,
	Amir~Jones Rafipoor, Mona Rafipoor, Klaus Sengstock, and Patrick
	Windpassinger.
	\newblock {Ultrastable, Zerodur-based optical benches for quantum gas
		experiments.}
	\newblock {\em Applied optics}, 53(20):4468--74, 2014.
	
\end{thebibliography}

\end{document}